# Scattering-free routing of surface plasmon polariton waves with optical null medium

Ali Abdolali*, Atefeh Ashrafian, Hooman Barati Sedeh, Mohammad Hosein Fakheri

Applied Electromagnetic Laboratory, School of Electrical Engineering, Iran University of Science and Technology, Tehran 1684613114, Iran

**Abstract:** Recently, guiding electromagnetic surface waves without sacrificing scattering losses through paths that have arbitrary shape bumps has gained a lot of interest due to its wealth of advantages in modern photonics and plasmonics devices. In this study, based on transformation optics (TO) methodology, a feasible approach to control the flow of surface plasmons polaritons (SPPs) at metal-dielectric interfaces with arbitrary curvature is proposed. The obtained material becomes homogeneous and independent of the bump's geometry. That is, one constant material is required to route SPP waves without scattering the energy into the far-field region, which overcome the bottlenecks encountered in the previous works. Several numerical simulations are carried out to illustrate the capability of the propounded cloak to control the SPP flows at metal/dielectric interfaces. The unique designing approach introduced here may open a new horizon to nano-optics and downscaling of photonic circuits.

## 1  Introduction

Manipulating Electromagnetic (EM) waves in an arbitrary manner is of utmost importance in various fields of science and engineering. Several methods have been proposed for this aim, which among all the propounded approaches, the recently developed technique of transformation optics (TO) has received a lot of attention due to its capability to precisely control and manipulate the flow of light [1]. As it was proposed by Pendry et al., the core idea of TO is to create a link between Maxwell's equations described in one coordinate system (named as virtual domain) and their counterparts in another predefined coordinate system (called as physical domain). The result of creating such an equivalence between these two coordinate systems is a mathematical expression that relates the constitutive parameters (i.e., permittivity and permeability) of the virtual space to the physical one [2]. The connection between the properties of these two media will consequently yield light to propagate along the desired path [3]. Following this unprecedented capability of manipulating EM waves, several devices that were deemed impossible to be obtained with conventional methods have been introduced such as invisibility cloaks [1,4-6], illusion systems [7,8], field concentrators [9,10], field rotators [11], beam steering application [12-15], radar absorber [16] and imaging application [17]. However, it should be noted that the capability of the coordinate transformation methodology is not restricted to the EM domain and could be extended to other physics such as acoustics [18, 19], thermodynamics [20, 21], water waves and matter waves [22].

Recently, much attention has been paid to manipulate near field waves such as surface plasmon polaritons (SPPs) with the aid of TO methodology. In particular, SPPs are kind of EM waves that are capable of propagating along the metal-dielectric interfaces. The electric field of SPP is significantly confined in the close vicinity of metal-dielectric interface while exponentially decreases in the direction perpendicular to the interface. Due to the intrinsic deep subwavelength behavior, large field enhancement and strong field confinement of SPPs, they have become common to be utilized in applications such as high-sensitivity chemical and biological sensing [23,24], photonic integrated circuit [25,26], and subwavelength imaging [27,28]. However, the main drawback of the devices that utilize SPPs as a means of transforming data is that they are incapable of guiding the SPP around an inevitable bump perfectly. That is, any minor abrupt discontinuity on the interface where the SPPs are propagating will cause severe scattering, which consequently result in performance degradation. Several attempts have been proposed both theoretically and experimentally to obviate the mentioned issue including utilizing V-grooves and slots [29-31], finite dielectric load [32] and structuring the metal surface [33]. More recently, Liu et al. utilized TO method to guide SPP path and cloak it from the surface discontinuities [34]. However, TO-based structures generally require anisotropic and spatially inhomogeneous materials, which limit their practical implementation in real-life scenarios. To eliminate the anisotropy circumscrip-tion, quasi-conformal TO approach is established in [35]. Another way to simplify the required material is to apply linear transformation, which leads to solely homogeneous constitutive parameters [36]. However, all the aforementioned methods were limited to protrusions with predefined regular shapes. To mitigate this drawback, the authors in [37] utilized infinitely anisotropic metamaterials to guide SPPs through arbitrary distorted metal surfaces, including slopes, bumps, and sharp corners. However, using infinitely anisotropic metamaterial leads to impedance mismatch problem. To solve this issue, the authors assumed a background with higher permittivity value compared to free space, which makes this work impractical for real-life scenarios since in most practical cases the background is free space. As well as the realization procedure, previously proposed mapping functions and their corresponding materials are strictly dependent on the desired disorder shapes. That is, if the structure shape is changed, one must recalculate the necessitating materials and as a result, the realization procedure should be repeated which is impractical and time consuming. Therefore, according to the above-mentioned points, it could be understood that another added layer of complexity in this field of research would be the manipulation of SPP flow at metal/dielectric interfaces that have arbitrary curvature with a constant material that is independent of the surface geometry. Nevertheless, optical surface transformation (OST) methodology, which is a branch of TO, has recently gained lots of attentions due to its unprecedented capability of manipulating light path with simple feasible shape independent materials that are called optic null medium (ONM) [38]. In this paper by extending the idea of OST methodology, we will propose a design principal which leads to a geometric free SPP cloak that can transfer the information without any scattering loss on any arbitrary shape surface. Since, the obtained ONMs are homogeneous, they could be easily implemented via metamaterial technology. Several numerical simulations have been carried out to demonstrate the performance of the designed cloak. Since, the obtained materials in the presented approach are independent of the cloak shape, they could be reused for transferring information on other surfaces with different structure geometry. This will consequently makes a huge step towards reconfigurable cloaks which is of utmost importance in today's life.

## 2  Design principles

To control the propagation of SPPs at metal/dielectric interfaces using coordinate transformation, one should modify both metallic and dielectric region [39]. However, since the most energy predominantly resides inside the dielectric, it is reasonable to control SPPs by only modifying the dielectric material, while keeping the metal property constant. Fig. 1 (a) shows the transformation relation between the virtual space and the physical space above the metal/dielectric interface. To this aim, a slab with the width of $W_v$ is transformed into another slab with the width of $W_p$ with a mathematical expression given as

$$x' = \begin{cases} x & x' \in (-\infty, 0) \\ \dfrac{W_p}{W_v} x & x' \in [0, W_p) \\ x - W_v + W_p & x' \in [W_p, \infty) \end{cases} \quad (1)$$

The coordinate transformation function of Eq. (1), indicates that the thickness of the virtual space slab will be expanded by the ratio of $W_p/W_v$ in the physical space. On the other hand, according to transformation optics methodology, the demanding constitutive parameters of the physical space will then be obtained as $\varepsilon/\varepsilon_b = \mu/\mu_b = (\det(\Lambda))^{-1} \times \Lambda \times \Lambda^T$ where $\Lambda = \partial(x', y', z')/\partial(x, y, z)$ is the Jacobian matrix which relates the metrics of virtual space, $(x, y, z)$, to the ones of physical space, $(x', y', z')$. Therefore, by substituting Eq. (1) into the given relation, the necessitating materials of the physical space will be obtained as $\varepsilon/\varepsilon_b = \mu/\mu_b = \mathrm{diag}[W_p/W_v, W_v/W_p, W_v/W_p]$, in which $\mathrm{diag}[.]$ represents a diagonal matrix.

When the thickness of the slab in the virtual space approaches zero (i.e., $W_v \to 0$), its interfaces become significantly close to each other (see $A_1$ and $A_2$ in the left side of Fig. 1), that could be considered as a flat surface. Therefore, Eq. (1) indicates a function that maps a sheet into a region, which is known as null (nihility) transformation [40]. Thus, the demanding materials of the physical domain will be changed to

$$\frac{\overline{\overline{\varepsilon}}}{\varepsilon_b} = \frac{\overline{\overline{\mu}}}{\mu_b} = \begin{bmatrix} \infty & 0 & 0 \\ 0 & 0 & 0 \\ 0 & 0 & 0 \end{bmatrix} \quad (2)$$

which is known as ONM [40]. Hereafter, we will name the axis in which the constitutive parameters are extremely large along it as the main axis (the x-axis in Fig.1). In addition, the main axis of the ONM can be in any other directions which gives rises to route the propagation of SPPs in a desired manner.

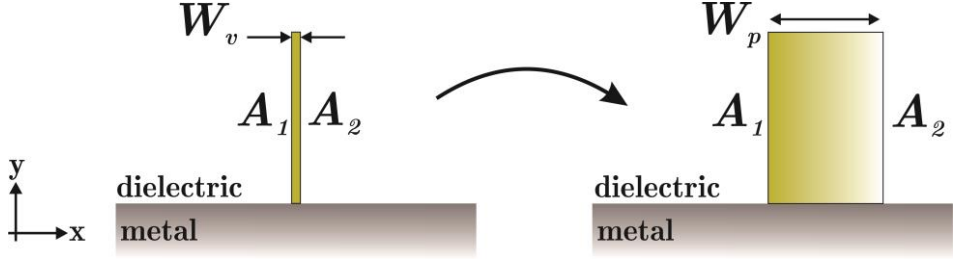

**Fig. 1:** The schematic of a coordinate transformation above the metal/dielectric interface that stretches a thin slab with thickness $W_v$ in the virtual space to a slab with finite thickness of $W_p$ in the physical space.

Moreover, the dispersion relation of a transverse magnetic (p-polarized) SPP wave propagating at the interface between a metal ( $z < 0$ ) and an anisotropic medium ( $z > 0$ ) can be written as [39]

$$\frac{k_{x_2}^2}{\varepsilon_z} + \frac{k_{z_2}^2}{\varepsilon_x} = \mu_y k_0^2 \quad (3)$$

which $k_{x_2}$ and $k_{z_2}$ are the longitudinal and transverse wave vector components in the anisotropic medium, respectively and $k_0$ is the free space wave number. Based on Eq. (3), since the value of approaches to infinity, regardless of the value of $k_{z_2}$ (in which the SPPs are evanescently distributed in the transverse direction), the SPPs constantly have the wave vector of $k_{x_2} = k_0\sqrt{\mu_y \varepsilon_z}$, that since $\mu_y = \varepsilon_z = 0$, the wave vector along the x-axis will become zero. Therefore, the fields of SPPs will be mapped point-to-point from one side to the other. In other words, an ONM will act as a void medium for the generated fields of SPP and thus they can transfer information without any phase distortion or accumulation from the input interface to its output one.

## 3 Simulations and results

To illustrate the capability of the propounded method in controlling the flow of SPPs on arbitrary shape surfaces, we have performed several numerical simulations with COMSOL MULTIPHYSICS finite-element solver. For the sake of simplicity, we have limited our discussions to two-dimensional (2D) scenario implying that our structure is unchanged in the z-direction. It should be remarked that the core idea and the underlying physics of ONM, is not restricted to the dimensions and could be easily extended to 3D case. For modelling the SPP propagation at the operative wavelength of $\lambda = 475\,\text{nm}$, silver has been utilized for the metal interface, which its dispersive permittivity follows the Drude model as $\varepsilon_m(\omega) = \varepsilon_\infty - \omega_p^2/(\omega^2 + i\gamma\omega)$, where $\varepsilon_\infty = 6$, $\omega_p = 1.5 \times 10^6\,\text{rad/s}$ and $\gamma = 7.7 \times 10^{13}\,\text{rad/s}$, and the dielectric region has been assumed to be filled with air. It is notable to mention that the height of the cloak must be higher than that of the decay length, which is related to the permittivity of metal as

$$\delta\big|_{\lambda_0 = 475\,\text{nm}} = \frac{1}{\text{Re}\left\{\sqrt{k_0/\left[-\varepsilon_m(\omega)\big|_{\lambda_0 = 475\,\text{nm}} - 1\right]}\right\}} = 0.2\,\mu\text{m} \quad (4)$$

The concept of SPP cloak based on utilization of ONM is shown schematically in Fig. 2(a). As it can be seen from this figure, each segment of the cloak is responsible for guiding the flow of SPP in the direction shown with red arrows. It should be mentioned that the materials that must be utilized for the sections tilted with the angle of $\alpha$, are obtained based on the multiplication of Eq. (2), when the deflection angle is zero, into the rotation matrix of $R(\alpha)$, which can be expressed as

$$\begin{bmatrix} \overline{\overline{\varepsilon}}_\alpha \\ \overline{\overline{\mu}}_\alpha \end{bmatrix} = R(\alpha) \times \begin{bmatrix} \overline{\overline{\varepsilon}} \\ \overline{\overline{\mu}} \end{bmatrix} \times R(\alpha)^T \tag{5}$$

In fact, by applying Eq. (5), which is a general form of Eq. (2), the SPP enters from surface $A_1$ and then without any phase accumulation and distortion it will be mapped point-to-point to interface $A_2$. In other words, these surfaces become equivalent and thus the flow of SPP can be smoothly guided around the bump without being distorted. It is noteworthy to mention that we have not performed any coordinate transformation in this case. In fact, the only thing that should be done in this method is to align the main axis of the ONM with the surface of the bump which is schematically shown in Fig. 2(a).

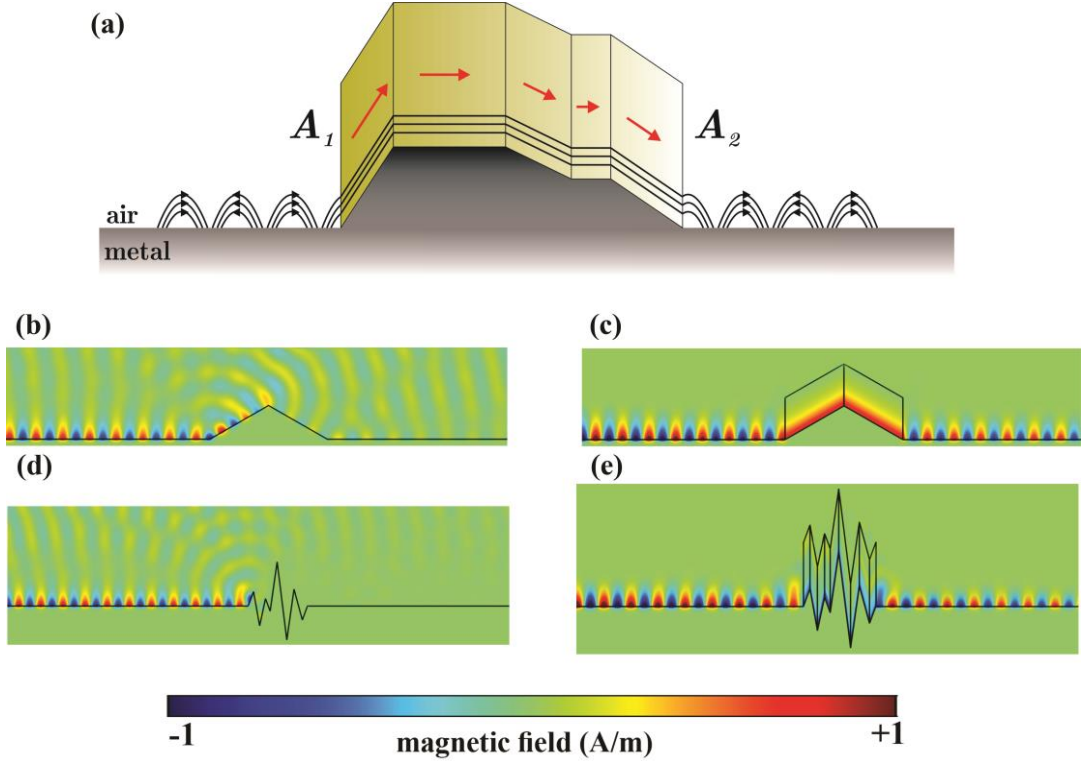

**Fig. 2:** The SPP guidance along the metallic bumps. (a) The schematic of the performance of the ONMs to guide the SPPs along an arbitrary multi-sections metallic bump. (b), (c) $H_z$-field distribution of a triangle-shape bump without and with the proposed material, respectively. (d), (e) $H_z$-field distribution of an arbitrary shape bump without and with the proposed material, respectively.

To reinforce the above mentioned points, we have studied the flow of SPP across a metallic triangle bump with $30^0$ leg angle with respect to the propagation direction (i.e. x-direction). When the cloak is not utilized, the SPP flows until the peak of the triangle bump and afterwards it will scatter into free space as it is shown numerically in Fig. 2(b), which yields the information loss. However, by utilizing the proposed method and exploiting the materials that are achieved based on Eq. (5) for $\alpha = \pm 30°$, the SPP flow can be smoothly guided around the triangular shape bump and thus lead to the data transportation with maximum efficiently as shown in Fig. 2(c). This kind of cloak can be also used for scenario when an arbitrary shape bump with multi-segments protrusion is locating in the path of SPPs wave. As it is shown in Fig. 2 (d), the SPP flow will be totally scattered and no data will be transferred to the other end of the bumps, when there is no cloak above the multiple bumps. However, utilizing the materials of Eq. (5) yields the impedance of each section to be matched with free space and thus guides the SPP flow without any scattering as it is demonstrated in Fig. 2(e). It should be remarked that although this kind of cloak might be achieved via performing linear transformation function proposed in [36], if the slope angle or the geometry of the bumps are changed, new materials must be recalculated and re designed which is impractical for realistic scenario. However, in the current work, we have only change the direction of the utilized materials along the predefined direction.

5In order to have full control on the flow of SPP, it is necessary to be able to manipulate SPPs on surfaces with irregular continuous bumps. Although several works based on quasi-conformal transformation optics (QCTO) have been performed for this aim [41], they are all geometry dependent, indicating that a change in the bump shape will lead to the recalculation of the demanding materials via tedious numerical approaches. Nevertheless, the competency of the presented ONM in this paper is not restricted only to segmental shape bumps and could be extended to other irregular geometries as well. To demonstrate this, we have first examined a semi-circle sin shape bump as shown in Fig. 3(a) when no cloak has been utilized for it. It is evident from the figure that due to scattering of SPP modes, modes profile will be distorted and thus yield the transferred data to be negligible.

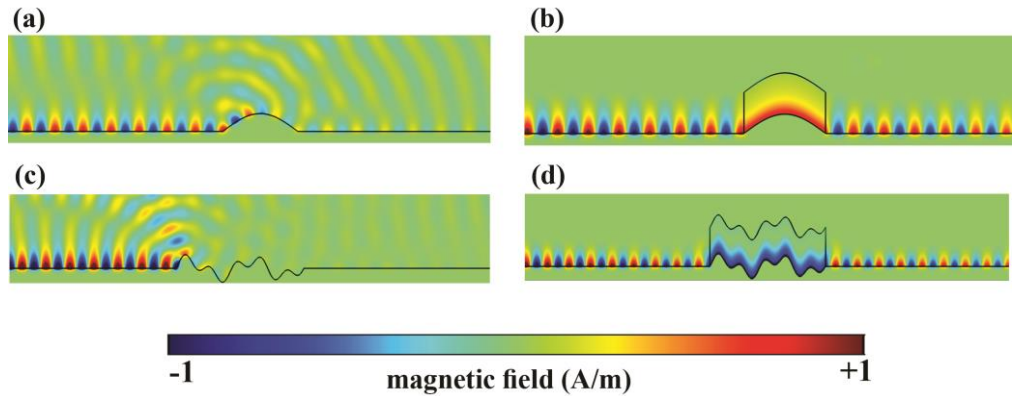

**Fig. 3:** The SPP guidance across the curvature metallic bumps. $H_z$-field distribution for (a), (b) bare and covered sine-shaped bump. (c), (d) bare and covered arbitrary shape bump.

Nevertheless, if the ONM main axis gradually varies with the obstacle slope, the information can be directly mapped from the input interface to the output one as it is shown in Fig. 3(b). It is clear that the SPP flow is smoothly guided around the bump indicating that the designed cloak well abides with the theoretical predictions. As the last example, we have assumed an arbitrary shape metallic bump with a curve function of $y = 0.4\sin(2x) - 0.3\cos(5x)$. As can be seen in Fig 4(e), without the ONM-based cloak, the SPPs scatter into the background due to abrupt change on the propagation surface and hence the fields after the curved bump is significantly negligible. However, by utilizing an ONM which its main axis is along the slope of the curvature, this scattering loss could be suppressed as shown in Fig. 4(d).

# 4  Conclusion

In conclusion, we have propounded a method to route the flow of SPP through metallic bumps with arbitrary shapes based on null-space transformation, which leads to a new material called ONM. The presented approach in this paper is capable of obviating inhomogeneity and shape dependency of conventional coordinate transformation-based approaches. That is, the designed ONM can be utilized for different arbitrary shape bumps and is competent to perfectly guide the SPP without any scattering. This will make the presented approach suitable for scenarios where reconfigurability is of utmost importance. To corroborate the effectiveness of our approach, several numerical simulations were performed for different shape surfaces. It was observed that the obtained results were well abide with theoretical predictions which validate the correctness of the present approach and verify the concept.